\documentclass[11pt]{article}
\usepackage{amsfonts}
\usepackage{amssymb,amsmath,mathrsfs}
\usepackage{graphicx}
\usepackage{subfigure}
\begin{document}
\title{\bf{A General Integrable Nonlocal Coupled Nonlinear Schr\"odinger Equation}}
\author{ Cai-Qin Song, Dong-Mei Xiao and Zuo-Nong Zhu
\footnote{Corresponding author. Email: znzhu@sjtu.edu.cn}
\\
Department of Mathematics, Shanghai Jiao Tong University,\\ 800 Dongchuan Road, Shanghai, 200240, P. R. China\\
}
\date{ }
\maketitle
\begin{abstract}
 In this paper, we investigate a general integrable nonlocal coupled nonlinear schr\"odinger (NLS) system with the the parity-time (PT) symmetry, which contains not only the nonlocal self-phase modulation and the nonlocal cross-phase modulation, but also the nonlocal four-wave mixing terms. This nonlocal coupled NLS system is a nonlocal version of a coupled NLS system. The general Nth Darboux transformation for the nonlocal coupled NLS equation is constructed. By using the Darboux transformation, its soliton solutions are obtained.
\end{abstract}
\section{Introduction}
\indent According to the work of Bender and Boettcher \cite {BB}, the PT symmetry plays a crucial role in the spectrum of the Hamiltonian. They proved that a wide class of non-Herimitan Hamiltons with PT symmetry have real and positive spectrum. This motivated the interest for many researchers on the PT symmetry in quantum mechanics [2-5]. For a non-Hermitian Hamiltonian $H=\frac{d^2}{dx^2}+V(x)$, if $V(x)$
satisfies $V(x)=V^*(-x)$, then it is PT symmetric. Under this condition, the Schr\"{o}dinger equation $i\Psi_t=H\Psi$ is PT symmetric. It has been shown that the optics can provide a good condition for testing the theory of PT symmetry or observing the phenomenon when the PT symmetry is breaking [6-9].\\
\indent As is well known, the nonlinear Schr\"{o}dinger equation (NLS),
\begin{eqnarray}
 iq_t(x,t)=q_{xx}(x,t)\pm 2|q(x,t)|^2q(x,t),
\end{eqnarray}
which is PT symmetric, received a wide study since the appearance of the work by Zakharov and Shabat \cite{ZS}. The NLS equation has many important physical applications, such as nonlinear optics \cite{GP}, water wave \cite {DJ}, and plasma physics. Very recently, Ablowitz
and Musslimani investigated a nonlocal NLS equation \cite {AM}:
\begin{equation}\label{nls}
  iq_t(x,t)=q_{xx}(x,t)\pm 2q(x,t)q^{*}(-x,t)q(x,t),
\end{equation}
which is derived from a new symmetry reduction of the well-known AKNS system,
where $q(x,t)$ is a complex valued function of the real variables $x$ and $t$ and $*$ denotes complex conjugation. The nonlocal NLS equation \eqref {nls} is an integrable system possessing the Lax pair, infinitely many conservation laws and it is solvable by using the inverse scattering transformation \cite {AM}.
Like the classical NLS equation,
the nonlocal NLS equation \eqref{nls} is PT symmetric and can be regarded as a mathematical model describing wave propagation phenomena
in PT symmetric nonlinear media. The nonlocal NLS equation possesses some new behaviors, e.g., it simultaneously admits both the bright and the dark soliton solutions \cite {AK}. In addition, the dark and antidark soliton interactions have been studied by use of the Darboux transformation method \cite {ML}.\\
On the other hand, integrable or nonintegrable coupled NLS equations have attracted widespread attention. In physical area, many problems, such as  the nonlinear light propagation in a birefringent optical fiber \cite{AH}, the two surface waves in deep water \cite{GJ},  Bose-Einstein condensates \cite{FD}, are governed by the coupled NLS equations.
In Ref. \cite{W}, integrable properties for the following general coupled NLS equation are investigated,
\begin{equation}\label{CNLS}
  \begin{array}{c}
  ip_t+p_{xx}+2[a|p|^2+c|q|^2+bpq^{*}+b^*qp^{*}]p=0,\\
  iq_t+q_{xx}+2[a|p|^2+c|q|^2+bpq^{*}+b^*qp^{*}]q=0,
  \end{array}
\end{equation}
which contains the cross-phase modulation, the self-phase modulation, and the four-wave mixing terms. When $a=c$ and $b=0$, equation \eqref{CNLS} reduces to Manakov equation \cite{M}. When $a=-c$ and $b=0$, it reduces to the mixed coupled NLS equation \cite {Z}. In \cite {W}, the N-soliton solutions in this system are obtained by the Riemann-Hilbert method. The collision of the two soltions in this system is analyzed. The dark soliton, breather and rogue wave of this system are given by the Hirota method and Darboux transformation \cite{N1,N2,N3}.\\
\indent In Ref.\cite {AS}, periodic and hyperbolic soliton solutions for a number of nonlocal nonlinear equations have been gained. Those nonlocal nonlinear equations include nonlocal NLS equation \eqref {nls} and the nonlocal coupled NLS equation,
\begin{eqnarray}\label{cnNLS}
&&ip_t(x,t)+p_{xx}(x,t)+[ap(x,t)p^{*}(-x,t)+b q(x,t)q^{*}(-x,t)]p(x,t)=0,\nonumber\\
&&iq_t(x,t)+q_{xx}(x,t)+[f p(x,t)p^{*}(-x,t)+e q(x,t)q^{*}(-x,t)]q(x,t)=0,
\end{eqnarray}
where $a,b,f,e$ are arbitrary real numbers.
In the case $a=b=e=f$ or $a=f=-e=-b$, this system reduces respectively to nonlocal Manakov equation or nonlocal Mikhailov-Zakharov-Schulman (MZS) equation. The authors of the Ref. \cite {AS} asked a question: whether like the nonlocal NLS equation, the coupled nonlocal NLS equation of Manakov type or MZS type is also an integrable system? They conjectured that indeed both the nonlocal Manakov and the nonlocal MZS are integrable systems.\\
In this paper, motivated by the investigation for the general coupled NLS equation \eqref{CNLS} and the nonlocal coupled NLS equation \eqref{cnNLS}, we consider the following general coupled nonlocal NLS equation:
\begin{eqnarray}\label{nonC}
&&ip_t+p_{xx}+2[app^{*}(-x,t)+cqq^{*}(-x,t)+bpq^{*}(-x,t)+b^*qp^{*}(-x,t)]p=0,\nonumber\\
&&iq_t+q_{xx}+2[app^{*}(-x,t)+cqq^{*}(-x,t)+bpq^{*}(-x,t)+b^*qp^{*}(-x,t)]q=0,
\end{eqnarray}
where $p=p(x,t), q=q(x,t)$ and the parameters $a$ and $c$ are real and $b$ is complex. It is obvious that equation \eqref{nonC} is a nonlocal version for the coupled NLS equation \eqref{CNLS}. The equation \eqref{nonC} contains the nonlocal cross-phase modulation, the nonlocal self-phase modulation, and the nonlocal four-wave mixing terms. In this paper, we will demonstrate the integrability for the coupled nonlocal NLS \eqref{nonC} by proposing its Lax pair. Inspired the work of the construction of Darboux transformation for the coupled NLS equation in Refs. \cite{L1,L2}, we will construct the Darboux transformation for the nonlocal coupled NLS equation \eqref{nonC}. As an application of the Darboux transformation, we will derive N-soliton solutions of the nonlocal coupled NLS equation \eqref{nonC}.

%The darboux transformation method[10] is one of the method to solve the integrable soliton problem,but in 1996[9],Manas give  dark soliton throught a new darboux transformation.In[6],Liming Ling,Lichen zhao and Boling Guo use this method to N-component coupled nonlinear Schrodinger equation and other type of coupled nonlinear Schrodinger equations[7],then get rich results, call it grneral darboux transformation.\\
%
%this problem were proposed by D.S.Wang,D.J.Zhang and J.K.Yang,they derive its N-bright soliton solutions by the Riemann-Hilbert method,
%and examine the collision dynamics between two solitons and relate the after-collision soliton
%parameters with the precollision ones.In[5],N. Vishnu Priya et give the dark-dark soliton, general breather, Akhmediev breather, Ma soliton, and rogue wave solutions,but we found the $p,q$ in those solution are linear related,and the bright-dark soliton is not given.
%This paper is arranged as following.In section 2, we give the binary darboux transformation of this problem and the N-soliton form.Insection 3,different types solution were got,Specially we got the breather-II solution,in which the p,q are not linear related.
\section{Darboux transformation for the coupled nonlocal NLS equation \eqref{nonC} }
In this section, we first demonstrate the integrability of the coupled nonlocal NLS \eqref{nonC}. Then we construct its Darboux transformation and soliton solutions. Inspired the form of the Lax pair for the coupled NLS equation \cite {W}, the Lax pair of equation \eqref{nonC} is given by
\begin{eqnarray}\label{lax}
  &&\Phi_x=U\Phi=(i\lambda\sigma_3+P)\Phi,\nonumber\\
  &&\Phi_t=V\Phi=-(2i\lambda^2\sigma_3+2\lambda P+i(P^2+P_x)\sigma_3)\Phi,
\end{eqnarray}
%\begin{equation}\label{uv}
%  \begin{array}{c}
%    U=\left(
%        \begin{array}{ccc}
%          i\lambda & p & q \\
%          -r_1 & -i\lambda & 0 \\
%          -r_2& 0 & -i\lambda \\
%        \end{array}
%      \right)
%    \\
%    \\
%    V=\left(
%        \begin{array}{ccc}
%          -2i\lambda^2+i(a|p|^2+c|q|^2+bpq^{*}+b^{*}p^{*}q )& -2\lambda p+ip_{x} & -2\lambda q+iq_{x} \\
%          2\lambda(ap^{*}+bq^{*})+i(ap_{x}^{*}+bq_{x}^{*}) & 2i\lambda^2-i(a|p|^2+bpq^{*})& -i(aqp^{*}+b|q|^{2}) \\
%          2\lambda(b^{*}p^{*}+cq^{*})+i(b^{*}p_{x}^{*}+cq_{x}^{*}) & -i(b^{*}|p|^2+cpq^{*}) & 2i\lambda^2-i(c|q|^2+b^{*}p^{*}q)\\
%        \end{array}
%      \right)
%  \end{array}
%\end{equation}
 where
\begin{equation*}
 P=\left(
     \begin{array}{ccc}
       0 & p & q\\
       -r_1& 0& 0\\
      -r_2 & 0& 0 \\
     \end{array}
   \right),
   \sigma_3=\left(
              \begin{array}{ccc}
                1& 0 & 0 \\
                0 &-1&0 \\
                0& 0 & -1 \\
              \end{array}
            \right),
\end{equation*}
and $r_1=a{p}^*(-x,t)+b {q}^*(-x,t),r_2={b^*}{p}^*(-x,t)+c {q^*}(-x,t)$. In fact, note that
$ \sigma_3P=-P \sigma_3$, and $ \sigma_3 P^2\sigma_3=P^2$ , we can check that the zero-curvature equation $U_t-V_x+[U,V]=0$ yields
\begin{eqnarray}
iP_t-P_{xx}\sigma_3+2P^3\sigma_3=0.
\end{eqnarray}
Equation (7) yields the coupled nonlocal NLS \eqref{nonC}. This implies the nonlocal coupled NLS equation \eqref{nonC} is an integrable system. Thus, both the nonlocal Manakov equation and the nonlocal MZS equation are integrable systems.\\
\indent Next, using a similar method of constructing the Darboux transformation for the coupled NLS equation \cite{L1,L2}, we construct the Darboux transformation for the coupled nonlocal NLS equation  \eqref{nonC}.
Suppose $\Phi=\Phi(x,t,\lambda),\Phi_1=\Phi(x,t,\lambda_1)$ are special vector solutions for system \eqref{lax} at $\lambda$ and $\lambda_1$ respectively. We define the binary function

%The binary Darboux transformation for coupled generalized nonlinear Schrodinger equations
%\begin{equation}
%T=I-\frac{\Phi_1 \Phi_{1}^{\dag} M}{i(\lambda-\bar{\lambda_1})\Omega(\Phi_1,\Phi_1)}
%\end{equation}
%here
\begin{equation}
 \Omega(\Phi_1,\Phi)=\frac{\Phi_{1}^{\dag}(-x,t)M \Phi}{i(\lambda+{\lambda_1^*})},
\end{equation}
where $\dag$ is the conjugate transpose, and $Re(\lambda_1)\not=0$. If $Re(\lambda_1)=0$, we choose $\Phi_1$ such that $\Phi_1$ satisfies $\Phi_{1}^{\dag}(-x,t)M \Phi_1=0$, and define  $\Omega(\Phi_1,\Phi_1)$ by
\begin{equation}\label{redt}
 \Omega(\Phi_1,\Phi_1)=\lim _{\lambda\rightarrow \lambda_1}\frac{\Phi_{1}^{\dag}(-x,t)M \Phi}{i(\lambda-\lambda_1)}+\theta,
\end{equation}
where $\theta$ is a proper constant, and
\begin{equation*}
  M=\left(
      \begin{array}{ccc}
        ac-|b|^2 & 0 & 0 \\
        0& -c& b\\
        0&b^{*} & -a \\
      \end{array}
    \right).
\end{equation*}
By a direct calculation, we have $\sigma_3M=M\sigma_3,P^{\dag}(-x,t)M=MP,P^{\dag}_x(-x,t)M=-MP_x,P\sigma_3=-\sigma_3P$. Then we can verify the following equations:
\begin{align*}
  \frac{d}{dx} \Phi^{\dag}(-x,t)M&=\Phi^{\dag}(-x,t)(i{\lambda^*}\sigma_3-P^{\dag}(-x,t))M=\Phi^{\dag}(-x,t)M(i{\lambda^*}\sigma_3-P),\\
   \frac{d}{dt} \Phi^{\dag}(-x,t)M &=\Phi^{\dag}(-x,t)\left(2i{\lambda}^{*2}\sigma_3-2{\lambda^*}P^{\dag}(-x,t)+i\sigma_3(P^{\dag 2}(-x,t)+P^{\dag}_x(-x,t))\right)M\\
   &= \Phi^{\dag}(-x,t)M\left(2i{\lambda}^{*2}\sigma_3-2{\lambda^*}P+i(P^2+P)\sigma_3\right).
\end{align*}
Thus, $\Omega(\Phi_1,\Phi)$ satisfies
\begin{align*}
  \Omega_x(\Phi_1,\Phi)&=\Phi_{1}^{\dag}(-x,t)M\sigma_3\Phi,\\
\Omega_t(\Phi_1,\Phi)&=2({\lambda^*}-\lambda)\Phi_{1}^{\dag}(-x,t)M\sigma_3\Phi-2\Phi_{1}^{\dag}(-x,t)MP\Phi.
\end{align*}
Introducing the following transformations for the eigenfunctions and the potentials:
\begin{equation}
\begin{array}{c}
\Phi[1]=\Phi-\displaystyle\frac{\Phi_1\Omega(\Phi_1,\Phi)}{\Omega(\Phi_1,\Phi_1)}, \\
  P[1]=P+[\sigma_3,\displaystyle\frac{\Phi_1\Phi_{1}^{\dag}(-x,t)M}{\Omega(\Phi_1,\Phi_1)}],
\end{array}
\end{equation}
then we can show
\begin{eqnarray}\label{DT}
&&\Phi[1]_x=(i\lambda\sigma_3+P[1])\Phi[1],\nonumber\\
&&\Phi[1]_t=-\left(2i\lambda^2\sigma_3+2\lambda P[1]+i(P^2[1]+P_x[1])\sigma_3\right)\Phi[1].
\end{eqnarray}
This means that the Darboux transformation for the nonlocal coupled NLS equation (5) is constructed. Here, we describe the main idea of the proof for equation \eqref{DT}.
Set $S=\displaystyle\frac{\Phi_1\Phi_{1}^{\dag}(-x,t)M}{\Omega(\Phi_1,\Phi_1)}$. A direct calculation yields
\begin{eqnarray*}
S^2=2i Re(\lambda_1)S,\qquad S\sigma_3S=S\displaystyle\frac{\Phi_{1}^{\dag}(-x,t)M\sigma_3\Phi_1}{\Omega(\Phi_1,\Phi_1)},\qquad SPS=S\displaystyle\frac{\Phi_{1}^{\dag}(-x,t)MP\Phi_1}{\Omega(\Phi_1,\Phi_1)}.
\end{eqnarray*}
By using those properties of the matrix $S$, we can verify the validity for equation \eqref{DT}.\\
Further, following the method of the constructing N-fold Darboux transformation for the coupled NLS in Refs. \cite{L1,L2}, we can obtain the N-fold Darboux transformation for the nonlocal coupled NLS equation \eqref{nonC}.\\
Suppose $\Phi_i$ (i=1,2...,N) are the N different eigenfunctions for the spectrum problem (6) at $\lambda=\lambda_i$ respectively, then the N-fold Darboux transformation can be written as
\begin{equation}%\label{}
 \Phi[N]=\Phi-RW^{-1}\Omega,
\end{equation}
where $R=(\Phi_1,\Phi_2,...,\Phi_N)$, and
\begin{equation}%\label{}
W=\left(
    \begin{array}{cccc}
      \Omega(\Phi_1,\Phi_1) & \Omega(\Phi_1,\Phi_2)  &\cdots &\Omega(\Phi_1,\Phi_N)  \\
      \Omega(\Phi_2,\Phi_1)  & \Omega(\Phi_2,\Phi_2)  & \cdots & \Omega(\Phi_2,\Phi_N)  \\
      \vdots& \vdots& \ddots & \vdots \\
     \Omega(\Phi_N,\Phi_1)  & \Omega(\Phi_N,\Phi_2)  &\cdots& \Omega(\Phi_N,\Phi_N) \\
    \end{array}
  \right),\Omega=\left(
                   \begin{array}{c}
                     \Omega(\Phi_1,\Phi) \\
                     \Omega(\Phi_2,\Phi) \\
                     \vdots \\
                    \Omega(\Phi_N,\Phi) \\
                   \end{array}
                 \right).
\end{equation}
The relation between new potential and the old one is given by
\begin{equation}\label{nDT}
 P[N]=P+[\sigma_3,RW^{-1}R^{\dag}(-x,t)M].
\end{equation}
Here we give the proof for equation $\Phi[N]_x=(i\lambda \sigma_3+P[N])\Phi[N]$. It is similar to show that the eigenfunction $\Phi[N]$ satisfies the time evolution equation. In fact, a direct calculation gives
\begin{align*}
  \Phi[N]_x&=(i\lambda \sigma_3+P)\Phi-R_x W^{-1}\Omega-RW^{-1}\Omega_x+RW^{-1}W_xW^{-1}\Omega\\
  &=(i\lambda \sigma_3+P)\Phi-(i\sigma_3R\Lambda+PR)W^{-1}\Omega-RW^{-1}R^{\dag}(-x,t)M\sigma_3\Phi+RW^{-1}R^{\dag}(-x,t)M\sigma_3RW^{-1}\Phi\\
  &=(i\lambda \sigma_3+P)\Phi-PRW^{-1}\Omega-RW^{-1}R^{\dag}(-x,t)M\sigma_3\Phi+RW^{-1}R^{\dag}(-x,t)M\sigma_3RW^{-1}\Phi \\
  & -i\sigma_3R\Lambda W^{-1}\Omega+\sigma_3RWR^{\dag}(-x,t)M\Phi-i\sigma_3RW^{-1}(\lambda I+{\Lambda^*} )\Omega\\
  &=(i\lambda \sigma_3+P)\Phi-PRW^{-1}\Omega-RW^{-1}R^{\dag}(-x,t)M\sigma_3\Phi+RW^{-1}R^{\dag}M\sigma_3RW^{-1}\Phi+\sigma_3RWR^{\dag}(-x,t)M\Phi\\
  & -i\lambda\sigma_3RW^{-1}\Omega-\sigma_3RW^{-1}(R^{\dag}(-x,t)MR-iW\Lambda)W^{-1}\Omega\\
  &=(i\lambda \sigma_3+P[N])\Phi[N],
\end{align*}
where $\Lambda=\text{diag}(\lambda_1,\lambda_2,\ldots,\lambda_N)$, and we have used the two identities $R^{\dag}(-x,t)MR=i(W\Lambda+{\Lambda^*}W)$,and $i(\lambda I+{\Lambda^*})\Omega-R^{\dag}(-x,t)M\Phi=0$.
\section{Soliton solution for the nonlocal coupled NLS \eqref{nonC} }
%\hspace*{1cm} In the following,to get different types of soliton solution ,we should use some proposition of matrix A, like negative or positive definiteness,so we firstly try to find a diagonal matrix A which is a congruence matrix of M.For different value of a,b and c ,the matrix A is diferent.
%\\
%A). $a=0,c=0,b\neq 0$.suppose
%%\begin{equation*}
%% Q=\left(
%%     \begin{array}{ccc}
%%      1 & 0 &0 \\
%%       0 &\frac{b}{|b|} & \frac{b}{|b|} \\
%%      0 & 1& -1 \\
%%     \end{array}
%%   \right)
%%\end{equation*}
%\begin{equation*}
% Q=\left(
%     \begin{array}{ccc}
%      1 & 0 &0 \\
%       0 &\frac{|b|}{2b} & \frac{1}{2} \\
%      0 & \frac{|b|}{2b} &  \frac{-1}{2}  \\
%     \end{array}
%   \right)
%\end{equation*}
%suppose $A=\text{diag}(-|b|^2,-2|b|,2|b|)$then we have
%\begin{equation*}
%M= Q^{\dag} AQ
%\end{equation*}
%B).$c\neq 0$,suppose
%\begin{equation*}
% Q=\left(
%     \begin{array}{ccc}
%      1 & 0 &0 \\
%       0 &1 & \frac{-b}{c} \\
%      0 &0&1 \\
%     \end{array}
%   \right)
%\end{equation*}
%and $A=\text{diag}(ac-|b|^2,c,a-\frac{|b|^2}{c})$then
%\begin{equation*}
%M=Q^{\dag}A Q
%\end{equation*}
%so it easy to ge that when $ac-|b|^2>0,c>0$,the matrix $M$ is positive definite.its doesn't exit $\Phi_1\neq 0$ satisfy $\Phi_1^{\dag} M\Phi_1 = 0$. \\
Suppose $Re(\lambda_1)\neq0$, and $\Phi_1=(f_1,f_2,f_3)^{T}$ is the eigenfunction for the spectral problem (6) corresponding to $\lambda=\lambda_1$, then the Darboux transformation yields
\begin{equation}
 \begin{array}{c}
   p[1]=p[0]-\displaystyle\frac{4iRe(\lambda_{1})}{\mathfrak{g}} \left(cf_{1}f_{2}^{*}(-x,t)-b^{*}f_{1}f_{3}^{*}(-x,t)\right),\\
   q[1]=q[0]-\displaystyle\frac{4iRe(\lambda_{1})}{\mathfrak{g}} \left(af_{1}f_{3}^{*}(-x,t)-bf_{1}f_{2}^{*}(-x,t)\right),
 \end{array}
\end{equation}
where $\mathfrak{g}=(ac-|b|^2)f_{1}f_{1}^{*}(-x,t)-cf_{2}f_{2}^{*}(-x,t)-af_{3}f_{3}^{*}(-x,t)+bf_{3}f_{2}^{*}(-x,t)+b^{*}f_{2}f_{3}^{*}(-x,t)$.\\
\indent Suppose $p=\rho_1e^{i(k_1x+\omega_1 t)},q=\rho_2e^{i(k_2 x+\omega_2 t)}$, where $\omega_i, k_i$ and $\rho_i$ are real numbers, are the seed solutions of nonlocal coupled NLS \eqref{nonC}, then $\omega_i$ and $k_i$ must satisfy
\begin{equation}
-\omega_i-k_i^2+2(a\rho_1^2e^{2ik_1x}+c\rho_2^2e^{2ik_2x}+b\rho_1\rho_2e^{i((k_1+k_2)x+(\omega_1-\omega_2)t)}+b^{*}\rho_1\rho_2e^{i((k_1+k_2)x+(\omega_2-\omega_1)t)})=0.
\end{equation}
This last equation yields $k_1=k_2=0$ and $\omega_1=\omega_2=\omega$. We thus see that $p=\rho_1e^{i\omega t}, q=\rho_2e^{i\omega t}$
are the seed solutions of nonlocal coupled NLS (5), where  $\omega=2m$, and $m=a\rho_1^2+c\rho_2^2+2b_R\rho_1\rho_2$. In order to solve spectral equation (6), we make the gauge transformation
\begin{equation*}
 \Phi=D \Psi ,  D=\text{diag}(1,e^{-i\omega t},e^{-i\omega t}),
\end{equation*}
then the spectral equation (6) leads to
\begin{equation}\label{Psi}
\Psi_x=i\tilde{U}\Psi,\Psi_t=i\tilde{V}\Psi,
\end{equation}
where
\begin{equation}
\begin{array}{c}
\tilde{U}=\left(
    \begin{array}{ccc}
      \lambda & -i\rho_1 & -i\rho_2 \\
      i(a\rho_1+b\rho_2) & -\lambda & 0 \\
     i(b^{*}\rho_1+c\rho_2) & 0 & -\lambda  \\
    \end{array}
     \right),\\
   \tilde{V}=\left(
        \begin{array}{ccc}
          -2\lambda^2+(a\rho_1^2+c\rho_2^2+2b_R\rho_1\rho_2) & i2\lambda\rho_1&  i2\lambda\rho_2 \\
           -i2\lambda(a\rho_1+b\rho_2)& 2\lambda^2-(a\rho_1^2+b\rho_1\rho_2)+\omega& -(a\rho_1\rho_2+b\rho_2^2) \\
         -i2\lambda(b^{*}\rho_1+c\rho_2) & -(b^{*}\rho_1^2+c\rho_1\rho_2)& 2\lambda^2-(c\rho_2^2+b^{*}\rho_1\rho_2)+\omega \\
        \end{array}
      \right).
      \end{array}
\end{equation}
Consider the characteristic equation of the matrix $\tilde{U}$:
\begin{equation*}
 det(\tau I-\tilde{U})=0,
\end{equation*}
%\begin{equation}\label{chara}
%\tau^2-ik\tau+(a\rho_1^2+c\rho_2^2+2b_R\rho_1\rho_2)+(\lambda-k)\lambda=0
%\end{equation}
i.e.,
\begin{equation}\label{chara}
(\tau+\lambda)(\tau^2-m-\lambda^2)=0.
\end{equation}
%\begin{equation}\label{cet}
% \gamma^2-i\omega \gamma+(a\rho_1^2+c\rho_2^2+2b_R\rho_1\rho_2)(k+2\lambda)^2+(\omega+2\lambda^2-(a\rho_1^2+c\rho_2^2+2b_R\rho_1\rho_2))(2\lambda^2-(a\rho_1^2+c\rho_2^2+2b_R\rho_1\rho_2))=0
%\end{equation}
It is obvious that the characteristic equation has three roots $-\lambda,\delta,-\delta$,
where $\delta=-\sqrt{\lambda^2+m}$.\\
%\begin{equation}
%\begin{array}{c}
%  \Delta_1=-(2\lambda-k)^2-4(a\rho_1^2+c\rho_2^2+2b_R\rho_1\rho_2),
% \Delta_2=(2\lambda+k)^2\Delta_1,\\
%  \delta_1=\sqrt{-\Delta_1},\delta_2=-(2\lambda+k)\delta_1=-\sqrt{-\Delta_2}.
%\end{array}
%\end{equation}
 %\Delta_2=(2\lambda+k)^2\Delta_1,\\
  %\delta=\sqrt{\Delta},%\delta_2=2\lambda\delta_1=-\sqrt{-\Delta_2}.
\textbf{Case 1}. The characteristic equation has three different roots and $\lambda_R\not=0$.\\
In this case, we obtain the eigenfunction $\Phi=(\phi_1,\phi_2,\phi_3)^T=D\Psi$, where $\Psi=HNv$, with $v=(1,\beta,\alpha)^{T}$, and
\begin{equation*}
N(x,t)=\text{diag}(e^{i(\delta x+(\frac{\omega}{2}-2\lambda\delta)t)},  e^{-i(\delta x-(\frac{\omega}{2}+2\lambda\delta)t)},e^{-i(\lambda x-(\omega+2\lambda^2)t)}),
\end{equation*}
\begin{equation*}
H=\left(
     \begin{array}{ccc}
    1  & 1 & 0 \\
      \displaystyle\frac{i(a\rho_1+b\rho_2)}{\delta+\lambda} & \displaystyle\frac{i(a\rho_1+b\rho_2)}{-\delta+\lambda}& -\rho_2 \\
        \displaystyle\frac{i(b^{*}\rho_1+c\rho_2)}{\delta+\lambda} &  \displaystyle\frac{i(b^{*}\rho_1+c\rho_2)}{-\delta+\lambda} & \rho_1\\
     \end{array}
   \right).
\end{equation*}
Further, we have
\begin{align*}
  \phi_1(\lambda,\delta)&=e^{i(\delta x+(\frac{\omega}{2}+2\lambda\delta)t)}+\beta e^{-i(\delta x-(\frac{\omega}{2}+2\lambda\delta)t)},\\
  \phi_2(\lambda,\delta)&=i(a\rho_1+b\rho_2)(\frac{e^{i(\delta x-(\frac{\omega}{2}+2\lambda\delta)t)}}{\lambda+\delta}+\beta\frac{e^{-i(\delta x+(\frac{\omega}{2}-2\lambda\delta)t)}}{\lambda-\delta})-\alpha \rho_2 e^{-i(\lambda x-2\lambda^2 t)},\\
  \phi_3(\lambda,\delta)&=i(b^{*}\rho_1+c\rho_2)(\frac{e^{i(\delta x-(\frac{\omega}{2}+2\lambda\delta)t)}}{\lambda+\delta}+\beta\frac{e^{-i(\delta x+(\frac{\omega}{2}-2\lambda\delta)t)}}{\lambda-\delta})+\alpha \rho_1 e^{-i(\lambda x-2\lambda^2 t)},
\end{align*}
and $\Phi_1=(f_1,f_2,f_3)^{T}=(\phi_1(\lambda_1,\delta_1),\phi_2(\lambda_1,\delta_1),\phi_3(\lambda_1,\delta_1))^T$.\\
Note that
$\Phi^{\dag}(-x,t)M \Phi=N^{\dag}(-x,t)H^{\dag}D^{\dag} M DHN=N^{\dag}(-x,t)H^{\dag} M HN$.
Set  $B=(B_{ij})_{3\times 3}=H^{\dag}M H$, then we have
\begin{equation*}
 B=(ac-|b|^2)\left(
               \begin{array}{ccc}
                 \displaystyle\frac{2\lambda_R}{\lambda_R+\delta_R} & \displaystyle\frac{2\lambda_R}{\lambda_R-i\delta_I}& 0 \\
                \displaystyle\frac{2\lambda_R}{\lambda_R+i\delta_I} & \displaystyle\frac{2\lambda_R}{\lambda_R-\delta_R} &0\\
                 0 &0 & \displaystyle\frac{m}{|b|^2-ac}\\
               \end{array}
             \right)
\end{equation*}
%\begin{equation*}
%  p[1]=\rho_1e^{i\omega t}(1-\frac{2i\lambda(\frac{i(|b|^2-ac)\rho_1}{\lambda+i\kappa}e^{4\lambda\kappa t}-\bar{\alpha}(c\rho_2+b^{*}\rho_1)e^{k(x-2\lambda t)-i(\lambda x+(2\lambda^2-\frac{\omega}{2})t)})}{2(ac-|b|^2)e^{4\lambda \kappa t}-(a\rho_1^2+c\rho_2^2+2b_R\rho_1\rho_2)|\alpha|^2e^{-2i\lambda x}})
%\end{equation*}
%\begin{equation*}
%  q[1]=\rho_2e^{i\omega t}(1-\frac{2i\lambda(\frac{i(|b|^2-ac)\rho_2}{\lambda+i\kappa}e^{4\lambda\kappa t}+\bar{\alpha}(b\rho_2+a\rho_1)e^{k(x-2\lambda t)-i(\lambda x+(2\lambda^2-\frac{\omega}{2})t)})}{2(ac-|b|^2)e^{4\lambda \kappa t}-(a\rho_1^2+c\rho_2^2+2b_R\rho_1\rho_2)|\alpha|^2e^{-2i\lambda x}})
%\end{equation*}
%
%when $v=(1,\beta,0)^{T}$,then
%\begin{equation*}
% p[1]=\rho_1e^{i\omega t}(1-\frac{4\lambda(e^{2\kappa(x+2\lambda t)}+\beta)(\frac{e^{-2\kappa(x-2\lambda t)}}{\lambda+i\kappa}+\frac{\bar{\beta}}{\lambda-i\kappa})}{2e^{8\lambda\kappa t}+\frac{2\beta\lambda e^{-2\kappa(x-2\lambda t)}}{\lambda+i\kappa}+\frac{2\bar{\beta}\lambda e^{2\kappa(x+2\lambda \kappa t)}}{\lambda-i\kappa}+2|\beta|^2})
%\end{equation*}
%and $q[1]=\frac{\rho_2}{\rho_1}p[1]$.\\
We thus obtain
\begin{align*}
   \Phi^{\dag}(-x,t)M \Phi&=B_{11}e^{i((\delta+\delta^*)x-2(\lambda\delta-\lambda^*\delta^*)t)}+\beta B_{12} e^{-i((\delta-\delta^*)x+2(\lambda\delta+\lambda^*\delta^*)t)}+\beta^* B_{21}e^{i((\delta-\delta^*)x-2(\lambda\delta+\lambda^*\delta^*)t)} \\
  & +|\beta|^2 B_{22}e^{-i((\delta+\delta^*)x-2(\lambda\delta-\lambda^*\delta^*)t)}+|\alpha|^2 B_{33}e^{-i((\lambda+\lambda^*)x-2(\lambda^2-\lambda^{*2})t)}.
\end{align*}\\
\indent Here we focus on the case of $\beta\not=0$, $\lambda_1\in {R}$, and $\lambda_1^2+m<0$. We can show that in this case $\Phi_1^{\dag}(-x,t)M \Phi_1\neq0$ for arbitrary $x$ and $t$. This means that nonsigular and local solution for nonlocal coupled NLS equation \eqref{nonC} can be obtained. We denote $\delta_1=-i\sqrt{-\lambda_1^2-m}=-i\kappa$,
then
\begin{eqnarray*}
\Phi_1^{\dag}(-x,t)M \Phi_1=2(ac-|b|^2)\mathfrak{R},
\end{eqnarray*}
where
\begin{eqnarray*}
\mathfrak{R}=e^{-4\lambda_1\kappa t}+\frac{\beta\lambda_1 e^{-2\kappa x}}{\lambda_1+i\kappa}+\frac{\beta^*\lambda_1 e^{2\kappa x}}{\lambda_1-i\kappa}+|\beta|^2 e^{4\lambda_1\kappa t}-\frac{m|\alpha|^2 e^{-2i\lambda_1 x}}{2(ac-|b|^2)}.
\end{eqnarray*}
In the following, for the sake of simplicity, we set $\lambda=\lambda_1$. By a direct calculation, we obtain
\begin{align}\label{sp1}
 p[1]=\rho_1e^{i\omega t}[1-\frac{2\lambda(e^{\kappa(x-2\lambda t)}+\beta e^{-\kappa(x-2\lambda t)})}{\mathfrak{R}}(\frac{e^{-\kappa(x+2\lambda t)}}{\lambda+i\kappa}+\frac{\beta^*e^{\kappa(x-2\lambda t)}}{\lambda-i\kappa}
-\frac{i\alpha^*(c\rho_2+b^*\rho_1)}{\rho_1(ac-|b|^2)}e^{-i(\lambda x+(2\lambda^2+\frac{\omega}{2}) t )})],\nonumber\\
\end{align}
\begin{align}\label{sq1}
 q[1]=\rho_2e^{i\omega t}[1-\frac{2\lambda(e^{\kappa(x-2\lambda t)}+\beta e^{-\kappa(x-2\lambda t)})}{\mathfrak{R}}(\frac{e^{-\kappa(x+2\lambda t)}}{\lambda+i\kappa}+\frac{\beta^*e^{\kappa(x+2\lambda t)}}{\lambda-i\kappa}
 +\frac{i\alpha^*(b\rho_2+a\rho_1)}{\rho_2(ac-|b|^2)}e^{-i(\lambda x+(2\lambda^2+\frac{\omega}{2}) t )})].
\end{align}
We remark here that if set $\alpha=0$, the solution yields the one of the nonlocal NLS equation, which has been given \cite{ML}.\\
Next we will analyse the asymptonic behavior of the solutions (\ref{sp1}) and (\ref{sq1}). We do this for solution $p[1]$, since the asymptonic analysis for the solution $q[1]$ is similar.\\
We rewrite the solution $p[1]$ as the following two different forms, respectively,
\begin{equation}
 p[1]=\rho_1e^{i\omega t}[1-\frac{2\lambda(1+\beta e^{-2\kappa(x-2\lambda t)})}{\mathfrak{R_1}}(\frac{ 1}{\lambda+i\kappa}+\frac{\beta^*e^{2\kappa(x+2\lambda t)}}{\lambda-i\kappa}-\frac{i\alpha^*(c\rho_2+b^*\rho_1)}{\rho_1(ac-|b|^2)}e^{-i(\lambda x+(2\lambda^2+\frac{\omega}{2}) t )+\kappa(x+2\lambda t)})],
\end{equation}
\begin{equation}
 p[1]=\rho_1e^{i\omega t}[1-\frac{2\lambda(e^{2\kappa(x-2\lambda t)}+\beta)}{\mathfrak{R_2}}(\frac{e^{-2\kappa(x+2\lambda t)}}{\lambda+i\kappa}+\frac{\beta^*}{\lambda-i\kappa}-\frac{i\alpha^*(c\rho_2+b^*\rho_1)}{\rho_1(ac-|b|^2)}e^{-i(\lambda x+(2\lambda^2+\frac{\omega}{2}) t)-\kappa(x+2\lambda t)})],
\end{equation}
where
\begin{align*}
      && \mathfrak{R_1}=1+\frac{\beta\lambda e^{-2\kappa (x-2\lambda t)}}{\lambda+i\kappa}+\frac{\beta^*\lambda e^{2\kappa (x+2\lambda t)}}{\lambda-i\kappa}+|\beta|^2 e^{8\lambda\kappa t}-\frac{m|\alpha|^2 e^{-2i\lambda x+4\lambda\kappa t}}{2(ac-|b|^2)}, \\
       &&\mathfrak{R_2}=e^{-8\lambda\kappa t}+\frac{\beta\lambda e^{-2(\kappa x+2\lambda t)}}{\lambda+i\kappa}+\frac{\beta^*\lambda e^{2\kappa (x-2\lambda t)}}{\lambda-i\kappa}+|\beta|^2 -\frac{m|\alpha|^2 e^{-2i\lambda x-4\kappa\lambda t}}{2(ac-|b|^2)}.
     \end{align*}\\
\indent (A) Along the line $x+2\lambda t=0$, we have
\begin{equation}\label{19}
\begin{split}
  p[1]&=\rho_1e^{i\omega t}\left(1-\frac{2\lambda(\frac{ 1}{\lambda+i\kappa}+\frac{\beta^*e^{2\kappa(x+2\lambda t)}}{\lambda-i\kappa}-\frac{i\alpha^*(c\rho_2+b^*\rho_1)}{\rho_1(ac-|b|^2)}e^{-i(\lambda x+(2\lambda^2+\frac{\omega}{2}) t )+\kappa(x+2\lambda t)})}
 {1+\frac{\beta*\lambda e^{2\kappa (x+2\lambda t)}}{\lambda-i\kappa}}\right)  \\
    &=\rho_1e^{i\omega t}\left(\frac{-\lambda}{\lambda+i\kappa}-\frac{i\kappa}{\lambda+i\kappa}tanh(\theta_1)+\frac{i\alpha^*(c\rho_2+b^*\rho_1)}{\rho_1(ac-|b|^2)}e^{-i(\lambda x+(2\lambda^2+\frac{\omega}{2}) t )-\varepsilon_1-i\zeta}sech(\theta_1)\right)
\end{split} \quad \lambda t \rightarrow -\infty
\end{equation}
%\begin{equation}
% p[1]=\rho_1e^{i\omega t}(1-\frac{\frac{ 1}{\lambda+i\kappa}+\frac{\bar{\beta}e^{2\kappa(x-2\lambda t)}}{\lambda-i\kappa}-\frac{i\bar{\alpha}(c\rho_2+\bar{b}\rho_1)}{\rho_1(ac-|b|^2)}e^{-i(\lambda x+(2\lambda^2+\frac{\omega}{2}) t )+\kappa(x-2\lambda t)}}
% {1+\frac{\bar{\beta}\lambda e^{2\kappa (x-2\lambda t)}}{\lambda-i\kappa}})     \qquad\quad \lambda t \rightarrow +\infty
%\end{equation}
\begin{equation}\label{20}
\begin{split}
 p[1]&=\rho_1e^{i\omega t}\left(1-\frac{2\lambda\beta(\frac{e^{-2\kappa(x+2\lambda t)}}{\lambda+i\kappa}+\frac{\beta^*}{\lambda-i\kappa}-\frac{i\alpha^*(c\rho_2+b^*\rho_1)}{\rho_1(ac-|b|^2)}e^{-i(\lambda x+(2\lambda^2+\frac{\omega}{2}) t)-\kappa(x+2\lambda t)})}{\frac{\beta\lambda e^{-2(\kappa x+2\lambda t)}}{\lambda+i\kappa}+|\beta|^2}\right) \\
    &=\rho_1e^{i\omega t}\left(\frac{-\lambda}{\lambda-i\kappa}-\frac{i\kappa}{\lambda-i\kappa}tanh(\theta_2)+\frac{2i\lambda\alpha^*(c\rho_2+b^*\rho_1)}{\beta^*\rho_1(ac-|b|^2)}e^{-i(\lambda x+(2\lambda^2+\frac{\omega}{2}) t )+\varepsilon_2-i\zeta}sech(\theta_2)\right)
\end{split}  \lambda t \rightarrow +\infty
\end{equation}
where
\begin{equation*}
\theta_1=\kappa(x+2\lambda t)+\varepsilon_1+i\zeta,2\varepsilon_1=ln(\frac{|\beta^*\lambda|}{|\lambda-i\kappa|}),2\zeta=Arg(\frac{ \beta^*\lambda}{\lambda-i\kappa});
\end{equation*},
\begin{equation*}
  \theta_2=\kappa(x+2\lambda t)+\varepsilon_2+i\zeta,-2\varepsilon_2=ln(\frac{|\lambda|}{|\beta^*(\lambda+i\kappa)|}).
\end{equation*}
\\
\indent(B) Along the line $x-2\lambda t=0$, we have

\begin{equation}\label{21}
%\begin{split}
  p[1]=\rho_1e^{i\omega t}\left(1-\frac{2\lambda(1+\beta e^{-2\kappa(x-2\lambda t)})}{\lambda+i\kappa+\beta \lambda e^{-2\kappa(x-2\lambda t)}}\right)
      =\rho_1e^{i\omega t}\left(\frac{-\lambda}{\lambda+i\kappa}+\frac{i\kappa}{\lambda+i\kappa}tanh(\eta_1)\right),\qquad\quad \lambda t \rightarrow -\infty
%\end{split}
\end{equation}
\begin{equation}\label{22}
%\begin{split}
  p[1]=\rho_1e^{i\omega t}\left(1-\frac{2\lambda( e^{2\kappa(x-2\lambda t)}+\beta)}{\lambda e^{2\kappa(x-2\lambda t)}+\beta(\lambda-i\kappa)}\right)=\rho_1e^{i\omega t}\left(\frac{-\lambda}{\lambda-i\kappa}+\frac{i\kappa}{\lambda-i\kappa}tanh(\eta_2)\right),\qquad\quad \quad\lambda t \rightarrow +\infty
%\end{split}
\end{equation}
where
\begin{eqnarray*}
&&\eta_1=\kappa(x-2\lambda t)+\mu_1+i\nu,-2\mu_1=ln(\frac{|\beta\lambda|}{|\lambda+i\kappa|}),-2\nu=Arg(\frac{\beta \lambda}{\lambda+i\kappa}),\\
&&\eta_2=\kappa(x-2\lambda t)+\mu_2+i\nu,2\mu_2=ln(\frac{|\lambda|}{|\beta(\lambda-i\kappa)|}).
\end{eqnarray*}
\indent We can see that the phase shifts along the line $x+2\lambda t=0$ and the line $x-2\lambda t=0$ are equal, since $|\varepsilon_1-\varepsilon_2|=|\mu_1-\mu_2|=ln(\frac{|\lambda+i\kappa|}{|\lambda|})$.
It follows from (\ref{21}) and (\ref{22}) that the extreme value of $|p[1]|$ is $|\rho_1|\frac{|\lambda+\kappa tan(\nu)|}{|\lambda+i\kappa|}$. So, when $\lambda sin(2\nu)>\kappa cos(2\nu)$, the solution $p[1]$ is an antidark soliton. Note that $sin(2\nu)=\frac{\lambda(\beta_R \kappa-\beta_I \lambda)}{|\lambda\beta(\lambda-i\kappa)|},cos(2\nu)=\frac{\lambda(\beta_R \lambda+\beta_I \kappa)}{|\lambda\beta(\lambda-i\kappa)|}$,  we can get the conclusion that the soliton along the line $x-2\lambda t=0$ is antidark when $\lambda\beta_I<0$, and is dark when $\lambda\beta_I>0$. Similarly, when $\alpha^*(c\rho_2+b^*\rho_1)=0$, along the line $x+2\lambda t=0$, the soliton is dark if $\lambda Im(\beta(\lambda-i\kappa)^2)<0$, and is antidark if $\lambda Im(\beta(\lambda-i\kappa)^2)>0$. Those phenomenons are the same as the solution for the nonlocal NLS equation. But, when $\alpha^*(c\rho_2+b^*\rho_1)\neq0$, there are some differences along the line $x+2\lambda t=0$. In fact, it follows from the equations (\ref{19}) and (\ref{20}) that the solution $p[1]$ is a combination of the bright and the dark soliton or the bright and the antidark soliton. Especially, when $\rho_1=0, {\alpha^*}(c\rho_2+{b^*}\rho_1)\neq0$, the solution $p[1]$ degenerates to a bright soliton with a phase shift. The solutions of the nonlocal coupled NLS equation (5) are more richer than the ones of the nonlocal NLS equation.\\
\indent \textbf{Case 2}. $Re(\lambda_1)=0$ and $a\rho_1^2+c\rho_2^2+2b_R\rho_1\rho_2+\lambda_1^2>0$.\\
In this case, to obtain the solution $p[1]$ and $q[1]$, we must carefully determine $\Omega(\Phi_1, \Phi_1)$.
Choosing
\begin{equation}
\Phi=D\left(
          \begin{array}{c}
            1 \\
            \displaystyle\frac{i(a\rho_1+b\rho_2)}{\delta+\lambda} \\
            \displaystyle\frac{i(b^{*}\rho_1+c\rho_2)}{\delta+\lambda} \\
          \end{array}
        \right)e^{i(\delta x+(\frac{\omega}{2}-2\lambda \delta )t)}=\left(
                                                                         \begin{array}{c}
                                                                           e^{i(\delta x+(\frac{\omega}{2}-2\lambda \delta )t}) \\
                                                                           \displaystyle\frac{i(a\rho_1+b\rho_2)}{\delta +\lambda }e^{i(\delta x-(\frac{\omega}{2}+2\lambda\delta)t)} \\
                                                                           \displaystyle\frac{i(b^{*}\rho_1+c\rho_2)}{\delta+\lambda}e^{i(\delta x-(\frac{\omega}{2}+2\lambda \delta )t)} \\
                                                                         \end{array}
                                                                       \right),
\end{equation}
then we have
\begin{eqnarray}
  \Phi_1^{\dag}(-x,t)M\Phi=(ac-|b|^2)\left(1-\frac{a\rho_1^2+c\rho_2^2+2b_R\rho_1\rho_2}{(\lambda+\delta)(\delta_1-\lambda_1)}\right)e^{i((\delta+\delta_1)x-(2\lambda\delta+2\lambda_1\delta_1)t)}.
\end{eqnarray}
Note that, for the different spectral $\lambda$ and $\lambda_1$, we have $\lambda^2-\delta^2=\lambda_1^2-\delta_1^2$, and
\begin{equation*}
\frac{a\rho_1^2+c\rho_2^2+2b_R\rho_1\rho_2}{(\lambda+\delta)(\delta_1-\lambda_1)}-\frac{\lambda_1+\delta_1}{\lambda+\delta}=0.
\end{equation*}
We thus obtain
\begin{eqnarray}
 \Phi_1^{\dag}(-x,t)M\Phi=\frac{(\lambda-\lambda_1)(\lambda+\delta+\lambda_1+\delta_1)(ac-|b|^2)}{(\lambda+\delta)(\delta+\delta_1)}e^{i((\delta+\delta_1)x-(2\lambda\delta+2\lambda_1\delta_1)t)}.
\end{eqnarray}
This means that $\Phi_1^{\dag}(-x,t)M\Phi_1=0$, and $ \displaystyle \lim_{\lambda \rightarrow \lambda_1}\frac{\Phi_1^{\dag}(-x,t)M\Phi}{i(\lambda-\lambda_1)}=\frac{i(|b|^2-ac)}{\delta_1}e^{2i\delta_1(x-2\lambda_1 t)}$. Finally,  $\Omega(\Phi_1,\Phi_1)$ is determined by
\begin{equation}
 \Omega(\Phi_1,\Phi_1)=\frac{i(|b|^2-ac)}{\delta_1}e^{2i\delta_1x+4\delta_1k_1t}+\theta,
\end{equation}
where $\theta=\frac{i(|b|^2-ac)}{\delta_1}, \lambda_1=ik_1$. The solutions $p[1]$ and $q[1]$ are given by
\begin{eqnarray*}
&&p[1]=\rho_1e^{i\omega t}\left(\frac{\delta_1+\lambda_1}{\lambda_1-\delta_1}+\frac{2\delta_1}{\delta_1-\lambda_1}\frac{1}{e^{2i\delta_1x+4\delta_1k_1 t)}+1}\right),\nonumber\\
&&q[1]=\rho_2e^{i\omega t}\left(\frac{\delta_1+\lambda_1}{\lambda_1-\delta_1}+\frac{2\delta_1}{\delta_1-\lambda_1}\frac{1}{e^{2i\delta_1x+4\delta_1k_1 t)}+1}\right).\\
\end{eqnarray*}
We remark here that the solution is singular at $(x,t)=(\frac{(2k+1) \pi}{2 \delta_1}, 0), k\in \mathbb{N}$.\\
\indent \textbf{Case 3}. The characteristic equation has two multiple roots $\delta=0$.\\
In this case, let $m<0$, then $\lambda_1=\sqrt{-m}$. We obtain the eigenfunction $\Phi_1=(f_1,f_2,f_3)^T$, where
\begin{align*}
  f_1&=(1+\beta(x-2\lambda_1 t)) e^{\frac{i\omega t}{2}}, \\
   f_2&=\frac{i(a\rho_1+b\rho_2)}{\lambda_1}\left(1+\frac{\beta i}{\lambda_1}+\beta(x-2\lambda_1 t)\right)e^{\frac{-i\omega}{2}t}-\alpha \rho_2 e^{i(-\lambda_1 x+2\lambda_1^2t)},\\
   f_3&=\frac{i(b^{*}\rho_1+c\rho_2)}{\lambda_1}\left(1+\frac{\beta i}{\lambda_1}+\beta(x-2\lambda_1 t)\right)e^{\frac{-i\omega}{2}t}+\alpha \rho_1 e^{i(-\lambda_1 x+2\lambda_1^2t)}.
\end{align*}
Thus, solutions $p[1]$ and $q[1]$ of the nonlocal coupled NLS equation are given by
\begin{align*}
   p[1]&=\rho_1e^{i\omega t}\left(1-\frac{4}{\mathfrak{Y}}(\beta(x-2\lambda_1 t)+
   1)(1-\frac{i\beta^*}{\lambda_1}-\beta^*(x+2\lambda_1 t)+\frac{\lambda_1(c\rho_2+b^*\rho_1)\alpha^*}{i(ac-|b|^2)\rho_1}e^{-i(\lambda_1 x+(2\lambda_1^2+\frac{\omega}{2})t)})\right),\\
   q[1]&=\rho_2e^{i\omega t}\left(1-\frac{4}{\mathfrak{Y}}(\beta(x-2\lambda_1 t)+1)(1-\frac{i\beta^*}{\lambda_1}-\beta^*(x+2\lambda_1 t)-\frac{\lambda_1(b\rho_2+a\rho_1)\alpha^*}{i(ac-|b|^2)\rho_2}e^{-i(\lambda_1 x+(2\lambda_1^2+\frac{\omega}{2})t)})\right),
\end{align*}
where $\mathfrak{Y}=(1 -\beta^*(x+2\lambda_1 t))(1 +\beta(x-2\lambda_1 t))+(1-\frac{i \beta^*}{\lambda_1}-\beta^*(x+2\lambda_1 t))(1+\frac{i \beta}{\lambda_1}+\beta(x-2\lambda_1 t))-\frac{|\alpha|^2 m}{(ac-|b|^2)} e^{-2i\lambda_1 x}$. The case of $\alpha=0$ yields the rational solution of the nonlocal NLS equation (5).

\section{2-soliton solutions}
By using N-fold Darboux transformation, we can obtain the general wave solution for nonlocal coupled NLS equation \eqref {nonC}.
Suppose $\Phi_i=(f_{1i},f_{2i},f_{3i})^{T}$ (i=1,2...,N) are the N different eigenfunctions for the spectrum problem (6) at $\lambda=\lambda_i$ corresponding to
the nonzero background $p[0]=\rho_1e^{i\omega t},q[0]=\rho_2e^{i\omega t}$. N-soliton solution is given by
\begin{equation}\label{twosol}
  \begin{array}{c}
    p[N]=\rho_1e^{i\omega t}-2F_1W^{-1}(c F_2^{\dag}(-x,t)-b^{*}F_3^{\dag}(-x,t)), \\
     q[N]=\rho_2e^{i\omega t}-2F_1W^{-1}(aF_3^{\dag}(-x,t)-b F_2^{\dag}(-x,t)),
  \end{array}
\end{equation}
where
\begin{equation*}
 \begin{array}{c}
    F_1=(f_{11},f_{12},\cdots,f_{1N}), \\
    F_2=(f_{21},f_{22},\cdots,f_{2N}),\\
    F_3=(f_{31},f_{32},\cdots,f_{3N}).
   \end{array}
\end{equation*}
By use of the relation of $\Phi_i=D\Psi_i, D=diag (1, e^{-i\omega t}, e^{-i\omega t})$,  we have the determinant representation of the solution
\begin{equation}\label{nsol}
  \begin{array}{c}
    p[N]=\rho_1e^{i\omega t}\frac{det(W/2-X)}{det(W/2)},X=(X_{ij})_{N\times N}=e^{-i\omega t}(c F_2^{\dag}(-x,t)-b^{*}F_3^{\dag}(-x,t))F_1/\rho_1,\\
     q[N]=\rho_2e^{i\omega t}
     \frac{det(W/2-Y)}{det(W/2)},Y=(Y_{ij})_{N\times N}=e^{-i\omega t}(a F_3^{\dag}(-x,t)-b F_2^{\dag}(-x,t))F_1/\rho_2.
  \end{array}
\end{equation}
Let us give the case of $N=2$ in detail. Set the spectral parameters $\lambda_1$ and $\lambda_2$ are two real numbers. Then we have $\Phi_k=(f_{1k},f_{2k},f_{3k})^{T}, k=1,2$, where
\begin{eqnarray*}
&&f_{1k}=e^{i(\delta_k x+(\frac{\omega}{2}-2\lambda_k\delta_k)t)}+\beta_k e^{-i(\delta_k x-(\frac{\omega}{2}+2\lambda_k\delta_k)t)},\\
&&f_{2k}=\displaystyle\frac{i(a\rho_1+b\rho_2)}{\delta_k+\lambda_k} e^{i(\delta_k x-(\frac{\omega}{2}+2\lambda_k\delta_k)t)}+ \beta_k\displaystyle\frac{i(a\rho_1+b\rho_2)}{\lambda_k-\delta_k} e^{-i(\delta_k x+(\frac{\omega}{2}-2\lambda_k\delta_k)t)}-\alpha_k\rho_2 e^{-i(\lambda_k x-2\lambda_k^2 t)},\\
&&f_{3k}=\displaystyle\frac{i(b^{*}\rho_1+c\rho_2)}{\delta_k+\lambda_k} e^{i(\delta_k x-(\frac{\omega}{2}+2\lambda_k\delta_k)t)} + \beta_k\displaystyle\frac{i(b^{*}\rho_1+c\rho_2)}{\lambda_k-\delta_k} e^{-i(\delta_k x+(\frac{\omega}{2}-2\lambda_k\delta_k)t)}+\alpha_k\rho_1 e^{-i(\lambda_kx-2\lambda_k^2t)}.
\end{eqnarray*}
\indent {\bf Case 1} $\lambda_k^2+m<0, k=1,2$.\\
In this case, $\delta_k=-i\sqrt{-\lambda_k^2-m}=-i\kappa_k$. We thus have
\begin{align*}
\Omega(\Phi_1,\Phi_1)  &=i(|b|^2-ac)(\frac{e^{-4\lambda_1\kappa_1t}}{\lambda_1}+\frac{\beta_1e^{-2\kappa_1 x}}{\lambda_1+i\kappa_1}
   +\frac{\beta^*_1e^{2\kappa_1x}}{\lambda_1-i\kappa_1}+\frac{|\beta_1|^2e^{4\lambda_1\kappa_1 t}}{\lambda_1})+i\frac{m|\alpha_1|^2}{2\lambda_1}e^{-2i\lambda_1 x};\\
   \Omega(\Phi_1,\Phi_2)  &=2i(|b|^2-ac)(\frac{e^{(\kappa_2-\kappa_1)x-2(\lambda_2\kappa_2+\lambda_1\kappa_1)t}}{\lambda_1+\lambda_2+i(\kappa_1-\kappa_2)}
   +\frac{\beta_2e^{-(\kappa_2+\kappa_1)x-2(\lambda_1\kappa_1-\lambda_2\kappa_2)t}}{\lambda_1+\lambda_2+i(\kappa_1+\kappa_2)}
   \\
   &+\frac{\beta^*_1e^{(\kappa_2+\kappa_1)x-2(\lambda_2\kappa_2-\lambda_1\kappa_1)t}}{\lambda_1+\lambda_2-i(\kappa_1+\kappa_2)}
   +\frac{\beta^*_1\beta_2e^{(\kappa_1-\kappa_2)x+2(\lambda_1\kappa_1+\lambda_2\kappa_2)t}}{\lambda_1+\lambda_2-i(\kappa_1-\kappa_2)})+
   i\frac{m\alpha_2\alpha^*_1}{\lambda_1+\lambda_2}e^{i(-(\lambda_1+\lambda_2)x-2(\lambda_2^2-\lambda_1^2)t)};\\
\Omega(\Phi_2,\Phi_1)  &=2i(|b|^2-ac)(\frac{e^{(\kappa_1-\kappa_2)x-2(\lambda_2\kappa_2+\lambda_1\kappa_1)t}}{\lambda_1+\lambda_2+i(\kappa_2-\kappa_1)}
+\frac{\beta_2e^{-(\kappa_2+\kappa_1)x-2(\lambda_2\kappa_2-\lambda_1\kappa_1)t}}{\lambda_1+\lambda_2+i(\kappa_1+\kappa_2)}
   \\
   &+\frac{\beta^*_2e^{(\kappa_2+\kappa_1)x-2(\lambda_1\kappa_1-\lambda_2\kappa_2)t}}{(\lambda_1+\lambda_2)-i(\kappa_1+\kappa_2)}
   +\frac{\beta^*_2\beta_1e^{(\kappa_2-\kappa_1)x+2(\lambda_1\kappa_1+\lambda_2\kappa_2)t}}{\lambda_1+\lambda_2-i(\kappa_2-\kappa_1)})
   +i\frac{m\alpha_1\alpha^*_2}{\lambda_1+\lambda_2}e^{-i((\lambda_1+\lambda_2)x+2(\lambda_1^2-\lambda_2^2)t)};\\
\Omega(\Phi_2,\Phi_2)  &=i(|b|^2-ac)(\frac{e^{-4\lambda_2\kappa_2t}}{\lambda_2}+\frac{\beta_2e^{-2\kappa_2 x}}{\lambda_2+i\kappa_2}
+\frac{\beta^*_2e^{2\kappa_2x}}{\lambda_2-i\kappa_2}+\frac{|\beta_2|^2e^{4\lambda_2\kappa_2 t }}{\lambda_2})+i\frac{m|\alpha_2|^2}{2\lambda_2}e^{-2i\lambda_2 x};
\end{align*}
and
\begin{align*}
X_{11}&=i(|b|^2-ac)(e^{\kappa_1(x-2\lambda_1 t)}+\beta_1e^{-\kappa_1(x-2\lambda_1 t)})(\frac{e^{-\kappa_1(x+2\lambda_1 t)}}{\lambda_1+i\kappa_1}+\frac{e^{\kappa_1(x+2\lambda_1 t)}}{\lambda_1-i\kappa_1}-\frac{i\alpha^*_1(c\rho_2+b^*\rho_1)e^{-i(\lambda_1 x+2 \lambda_1^2t+\frac{\omega}{2} t)}}{\rho_1(ac-|b|^2)})\\
X_{12}&=i(|b|^2-ac)(e^{\kappa_2(x-2\lambda_2 t)}+\beta_2e^{-\kappa_2(x-2\lambda_2 t)})(\frac{e^{-\kappa_1(x+2\lambda_1 t)}}{\lambda_1+i\kappa_1}+\frac{e^{\kappa_1(x+2\lambda_1 t)}}{\lambda_1-i\kappa_1}-\frac{i\alpha^*_1(c\rho_2+b^*\rho_1)e^{-i(\lambda_1 x+2 \lambda_1^2t+\frac{\omega}{2} t)}}{\rho_1(ac-|b|^2)})\\
X_{21}&=i(|b|^2-ac)(e^{\kappa_1(x-2\lambda_1 t)}+\beta_1e^{-\kappa_1(x-2\lambda_1 t)})(\frac{e^{-\kappa_2(x+2\lambda_2 t)}}{\lambda_2+i\kappa_2}+\frac{e^{\kappa_2(x+2\lambda_2 t)}}{\lambda_2-i\kappa_2}-\frac{i\alpha^*_2(c\rho_2+b^*\rho_1)e^{-i(\lambda_2 x+2 \lambda_2^2t+\frac{\omega}{2} t)}}{\rho_1(ac-|b|^2)})\\
X_{22}&=i(|b|^2-ac)(e^{\kappa_2(x-2\lambda_2 t)}+\beta_2e^{-\kappa_2(x-2\lambda_2 t)})(\frac{e^{-\kappa_2(x+2\lambda_2 t)}}{\lambda_2+i\kappa_2}+\frac{e^{\kappa_2(x+2\lambda_2 t)}}{\lambda_2-i\kappa_2}-\frac{i\alpha^*_2(c\rho_2+b^*\rho_1)e^{-i(\lambda_2 x+2 \lambda_2^2t+\frac{\omega}{2} t)}}{\rho_1(ac-|b|^2)})\\
Y_{11}&=i(|b|^2-ac)(e^{\kappa_1(x-2\lambda_1 t)}+\beta_1e^{-\kappa_1(x-2\lambda_1 t)})(\frac{e^{-\kappa_1(x+2\lambda_1 t)}}{\lambda_1+i\kappa_1}+\frac{e^{\kappa_1(x+2\lambda_1 t)}}{\lambda_1-i\kappa_1}+\frac{i\alpha^*_1(b\rho_2+a\rho_1)e^{-i(\lambda_1 x+2 \lambda_1^2t+\frac{\omega}{2} t)}}{\rho_2(ac-|b|^2)})\\
Y_{12}&=i(|b|^2-ac)(e^{\kappa_2(x-2\lambda_2 t)}+\beta_2e^{-\kappa_2(x-2\lambda_2 t)})(\frac{e^{-\kappa_1(x+2\lambda_1 t)}}{\lambda_1+i\kappa_1}+\frac{e^{\kappa_1(x+2\lambda_1 t)}}{\lambda_1-i\kappa_1}+\frac{i\alpha^*_1(b\rho_2+a\rho_1)e^{-i(\lambda_1 x+2 \lambda_1^2t+\frac{\omega}{2} t)}}{\rho_2(ac-|b|^2)})\\
Y_{21}&=i(|b|^2-ac)(e^{\kappa_1(x-2\lambda_1 t)}+\beta_1e^{-\kappa_1(x-2\lambda_1 t)})(\frac{e^{-\kappa_2(x+2\lambda_2 t)}}{\lambda_2+i\kappa_2}+\frac{e^{\kappa_2(x+2\lambda_2 t)}}{\lambda_2-i\kappa_2}+\frac{i\alpha^*_2(b\rho_2+a\rho_1)e^{-i(\lambda_2 x+2 \lambda_2^2t+\frac{\omega}{2} t)}}{\rho_2(ac-|b|^2)})\\
Y_{22}&=i(|b|^2-ac)(e^{\kappa_2(x-2\lambda_2 t)}+\beta_2e^{-\kappa_2(x-2\lambda_2 t)})(\frac{e^{-\kappa_2(x+2\lambda_2 t)}}{\lambda_2+i\kappa_2}+\frac{e^{\kappa_2(x+2\lambda_2 t)}}{\lambda_2-i\kappa_2}+\frac{i\alpha^*_2(b\rho_2+a\rho_1)e^{-i(\lambda_2 x+2 \lambda_2^2t+\frac{\omega}{2} t)}}{\rho_2(ac-|b|^2)})
\end{align*}
{\bf Case 2}.  $m<0$, $\lambda_1^2+m<0$, and $\delta_1=-i\kappa_1$; $\delta_2=0$, and $\lambda_2=\sqrt{-m}$.\\
In this case, we have
\begin{align*}
 \Omega(\Phi_1,\Phi_2)&=\frac{(ac-|b|^2)}{i(\lambda_1+\lambda_2)}(1+\beta_2(x-2\lambda_2 t))(e^{-\kappa_1(x+2\lambda_1 t)}+\beta^*_1 e^{\kappa_1(x+2\lambda_1 t)})\\
   &-\frac{(ac-|b|^2)m}{i\lambda_2(\lambda_1+\lambda_2)}(\frac{i\beta_2}{\lambda_2}+\beta_2(x+2\lambda_2t)+1)(\frac{e^{-\kappa_1(x+2\lambda_1 t)}}{\lambda_1+i\kappa_1}+\frac{\beta^*_1e^{\kappa_1(x+2\lambda_1 t)}}{\lambda_1-i\kappa_1})\\
   &-\frac{\alpha_2\alpha_1^* m}{i(\lambda_1+\lambda_2)} e^{i(-(\lambda_1+\lambda_2)x+2(\lambda_2^2-\lambda_1^2)t)}, \\
\Omega(\Phi_2,\Phi_1)&=\frac{(ac-|b|^2)}{i(\lambda_1+\lambda_2)}(e^{\kappa_1(x-2\lambda_1 t)}+\beta_1 e^{-\kappa_1(x-2\lambda_1 t)})(1-\beta^*_2(x+2\lambda_2 t))\\
   &-\frac{(ac-|b|^2)m}{i\lambda_2(\lambda_1+\lambda_2)}(\frac{-i\beta^*_2}{\lambda_2}-\beta^*_2(x+2\lambda_2t)+1)(\frac{e^{\kappa_1(x-2\lambda_1 t)}}{\lambda_1-i\kappa_1}+\frac{\beta_1e^{-\kappa_1(x-2\lambda_1 t)}}{\lambda_1+i\kappa_1})\\
   &-\frac{\alpha^*_2\alpha_1 m}{i(\lambda_1+\lambda_2)} e^{i(-(\lambda_1+\lambda_2)x+2(\lambda_1^2-\lambda_2^2)t)}, \\
\Omega(\Phi_2,\Phi_2)  &=\frac{(ac-|b|^2)}{2i\lambda_2}(1+\beta_2(x-2\lambda_2 t))(1-\beta^*_2(x+2\lambda_2 t)\\
&-\frac{(ac-|b|^2)m}{2i\lambda_2^2}(1+\beta_2(x-2\lambda_2 t)+\frac{i\beta_2}{\lambda_2})(1-\beta^*_2(x+2\lambda_2 t)-\frac{i\beta^*_2}{\lambda_2})
-\frac{|\alpha_2|^2 m}{2i\lambda_2}e^{-2i\lambda_2 x},\\
X_{12}&=i(|b|^2-ac)(1+\beta_2(x-2\lambda_2 t) )\left(\frac{e^{-\kappa_1(x+2\lambda_1 t)}}{\lambda_1+i\kappa_1}+\frac{e^{\kappa_1(x+2\lambda_1 t)}}{\lambda_1-i\kappa_1}-\frac{i\alpha^*_1(c\rho_2+b^*\rho_1)e^{-i(\lambda_1 x+2 \lambda_1^2t+\frac{\omega}{2} t)}}{\rho_1(ac-|b|^2)}\right),\\
X_{21}&=i(|b|^2-ac)(e^{\kappa_1(x-2\lambda_1 t)}+\beta_1e^{-\kappa_1(x+2\lambda_1 t)})(\frac{-i\beta^*_2}{\lambda_2^2}\\
&-\frac{\beta^*_2}{\lambda_2}(x+2\lambda_2 t)+\frac{1}{\lambda_2}-\frac{i\alpha^*_2(c\rho_2+b^*\rho_1)e^{-i(\lambda_2 x+2 \lambda_2^2t+\frac{\omega}{2} t)}}{\rho_1(ac-|b|^2)}),\\
  X_{22} &=i(|b|^2-ac)(1+\beta_2(x-2\lambda_2 t)\left(\frac{-i\beta^*_2}{\lambda_2^2}-\frac{\beta^*_2}{\lambda_2}(x+2\lambda_2 t)+\frac{1}{\lambda_2}-\frac{i\alpha^*_2(c\rho_2+b^*\rho_1)e^{-i(\lambda_2 x+2 \lambda_2^2t+\frac{\omega}{2} t)}}{\rho_1(ac-|b|^2)}\right).
\end{align*}
Note that $ \Omega(\Phi_1,\Phi_1),X_{11},Y_{11}$ are same as the ones for the first case.
Thus, two-soliton solution is given by the formula \eqref{nsol} with $N=2$.
We remark here that if set $\alpha_k=0$, then $X=Y$, and thus the two-soliton solution of nonlocal coupled NLS (5) yields the one of the nonlocal NLS equation.\\

\section{Conclusions and Remarks}
In this paper, we have investigated a nonlocal coupled NLS equation. We have demonstrated its integrability by proposing its Lax pair. We have constructed its Nth Darboux transformation.
By using the Darboux transformation, we have derived its soliton solutions. The results obtained in this paper including the Lax pair, the Darboux transformation, and the soliton solutions can yield the corresponding ones for nonlocal NLS equation.
In the future, we will investigate the discrete version of the nonlocal coupled NLS equation (5).

\vskip 16pt \noindent {\bf
Acknowledgements} \vskip 12pt
The work of DMX is supported by the National Natural Science
Foundation of China (NSFC) under grants 11371248 and 11431008 and the RFDP of Higher Education of China under grant 20130073110074,
that of ZNZ by the NSFC under grant 11271254, and
in part by the Ministry of Economy and Competitiveness of Spain under
contract MTM2012-37070.

\small{

}
\end{document}